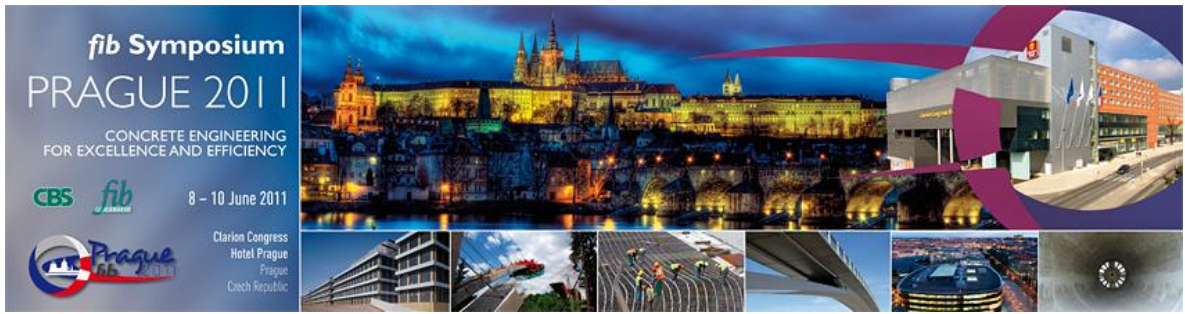

# ANALYSIS OF REINFORCED CONCRETE BEAMS BY THE EQUIVALENT SECTION METHOD

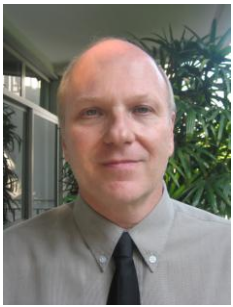 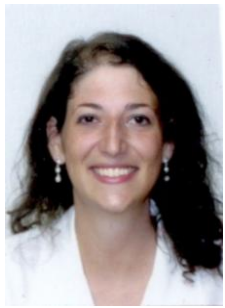

**Mauro Schulz**     **Maria Paola Santisi d'Avila**

## Summary

This research investigates the analysis of reinforced concrete beams subjected to combined axial load, bending moment and shear force. Cross-sections of general shape are divided, along the height, into plane elements. The biaxial behavior is represented according to the smeared rotating crack approach. Using traditionally accepted hypotheses for beams, the shear flow is determined by applying the Jouravski formula to an "equivalent section", which takes into account the nonlinear material behavior. The "Equivalent Section Method", originally proposed by Diaz (1980) and Diaz and Schulz (1981), is improved and simplified.

The formulation is implemented applying the bidimensional constitutive model A, proposed by Vecchio and Collins (1993). The tension-stiffening effect is considered as adopted by Polak and Vecchio (1993). Shear slip at crack surfaces, Poisson's ratio and other secondary effects are not considered. Validation is undertaken by comparison with experimental results obtained by other researchers. The examples include reinforced and prestressed concrete beams, for normal and high strength concrete. The formulation satisfactorily predicts the ultimate capacity under different load combinations. The whole set of equilibrium, compatibility and constitutive equations are satisfied, the stiffness derivatives are explicitly calculated and the algorithms show good convergence.

# 1　Introduction

The design of reinforced concrete members is traditionally based on different models for flexure and shear, even though shear forces interact with bending moments and axial loads.

Diaz (1980) proposed the concept of "equivalent section" and Diaz and Schulz (1981) developed a solution for reinforced concrete sections of general shape, submitted to axial load, bending moment and shear force. The method does not require the analysis of two close sections and the shear flow is determined by applying the Jouravski formula to the "equivalent section". Bentz (2000) presented a different numerical method that no longer requires the calculation of two sections, using other simplifying hypotheses.

The discussion here is limited to small displacements. As characteristic of a beam formulation, disturbed states of deformation at load application points and boundaries are not taken into account. The geometrical characteristics are considered constant along the beam and the stirrups lay in planes that are orthogonal to the longitudinal axis. The following hypotheses are introduced at the outset: 1. Eventual cracks are considered uniformly distributed and concrete stresses and strains are stated as continuous and derivable functions; 2. No slip is considered between concrete and reinforcing bars. Increments of steel and concrete strains are assumed to be equal on an average basis; 3. The beam can be analyzed according to a plane stress simplification; 4. The resultant of concrete and steel stresses in the transverse direction is small and, hence, neglected; 5. The principal directions of concrete stresses and strains are considered coincident; 6. Cross-sections remain plane after deformation; 7. In the interest of simplifying the formulation, normal forces are assumed to be constant along the beam and no forces are applied on the boundary of the cross-section.

# 2　Reinforced concrete element

Consider a beam with an arbitrary cross-section (**Fig. 1**). Tensile strains and stresses are positive. Concrete is assumed to be a continuous and uniform medium (hypothesis 1). The concrete strain vector $\boldsymbol{\varepsilon}$ at a level $z$ is defined by $\boldsymbol{\varepsilon} = \begin{bmatrix} \varepsilon_x & \varepsilon_z & \gamma_{xz} \end{bmatrix}^T$ where $\varepsilon_x$ and $\varepsilon_z$ are the normal strains in $x$- and $z$-directions. The shear strain is denoted by $\gamma_{xz}$.

According to hypothesis 2, the slip between steel bars and concrete is neglected. The stirrups lay in the $yz$-plane to simplify the formulation. However, the stirrup legs may be inclined to the $z$-axis at an angle $w$. The strains $\varepsilon_{sx}$ and $\varepsilon_{sw}$, respectively of the longitudinal and transverse reinforcements, are determined by $\varepsilon_{sx} = \varepsilon_x + \varepsilon_{sx0}$ and $\varepsilon_{sw} = \varepsilon_z \cos^2 w$. The strain $\varepsilon_{sx0}$ represents the residual strain of pre-tensioned or bonded post-tensioned tendons, which is calculated considering the tensioning operations, mobilized loading and prestressing losses.

The concrete stress vector at a level $z$ is $\boldsymbol{\sigma} = \begin{bmatrix} \sigma_x & \sigma_z & \tau_{xz} \end{bmatrix}^T$. It is assumed that the concrete stress components $\sigma_y$, $\tau_{yz}$ and $\tau_{xy}$ are approximately zero (hypothesis 3). The stress components $\sigma_x$, $\sigma_z$ and $\tau_{xz}$ at a point are independent of $y$, i.e., they do not vary through the thickness. The reinforcement stress vector $\boldsymbol{\sigma}_s$ is expressed by $\boldsymbol{\sigma}_s = \begin{bmatrix} \sigma_{sx} & \sigma_{sw} & 0 \end{bmatrix}^T$, where $\sigma_{sx}$ and $\sigma_{sw}$ are the stresses of the longitudinal and transverse reinforcements. The vector $\mathbf{f}$ of the axial forces $f_x$ and $f_z$ per unit length and shear flow $f_{xz}$ in the reinforced concrete element is equal to

$$\mathbf{f} = \begin{bmatrix} f_x & f_z & f_{xz} \end{bmatrix}^T = b\,\boldsymbol{\sigma} + \mathbf{a}_s\,\boldsymbol{\sigma}_s \tag{1}$$

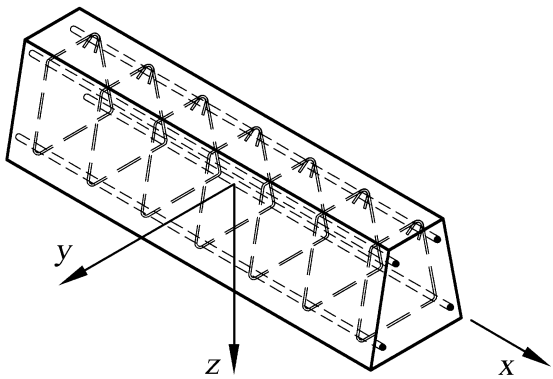 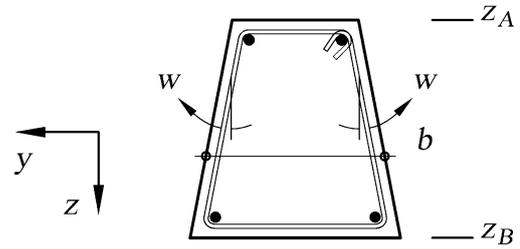

**Fig. 1** Coordinate system      **Fig. 2** Cross-section

As characteristic of a beam formulation, it is assumed $f_z \cong 0$ (hypothesis 4). The web thickness $b$ and steel area matrix $\mathbf{a}_s$ vary according to $z$-coordinate. The last is defined by

$$\mathbf{a}_s = \begin{bmatrix} a_{sx} & 0 & 0 \\ 0 & a_{sw}\cos w & 0 \\ 0 & 0 & 0 \end{bmatrix} \qquad (2)$$

where $a_{sx}$ is the area of the longitudinal reinforcement per unit $z$-length and $a_{sw}$ is the area of transverse reinforcement per unit $x$-length, considering the total number of legs. The reinforcement areas can be equal to zero.

Although several approaches for modeling the nonlinear behavior of reinforced concrete could be adopted, including path-dependent and irreversible processes, a simple hyperelastic stress-strain relationship is implemented. This procedure is usually considered simple and adequate for the limit analysis of reinforced concrete cross-sections under monotonic loadings. Assuming the coincidence of the principal directions (hypothesis 5), the concrete stress-strain relationship is expressed by $\boldsymbol{\sigma}_{12} = \boldsymbol{\sigma}_{12}(\boldsymbol{\varepsilon}_{12})$ where strain and stress vectors $\boldsymbol{\varepsilon}_{12}$ and $\boldsymbol{\sigma}_{12}$, in the principal coordinate system, are defined by $\boldsymbol{\varepsilon}_{12} = \begin{bmatrix} \varepsilon_1 & \varepsilon_2 & \gamma_{12} \end{bmatrix}^T$ and $\boldsymbol{\sigma}_{12} = \begin{bmatrix} \sigma_1 & \sigma_2 & \tau_{12} \end{bmatrix}^T$. Since shear stress $\tau_{12}$ and strain $\gamma_{12}$ are null in the principal coordinate system, the constitutive functions are replaced by:

$$\sigma_1 = \sigma_1(\varepsilon_1, \varepsilon_2) \qquad \sigma_2 = \sigma_2(\varepsilon_1, \varepsilon_2) \qquad (3)a,b$$

The angle $\theta$ is defined as the inclination of the principal compressive direction $x_2$ with respect to the $x$-axis The maximum and minimum principal strains, $\varepsilon_1$ and $\varepsilon_2$, as well as the principal direction $\theta$, are computed through standard compatibility relations.

Considering the coincidence of the principal directions, concrete strains and stresses are transformed according to the following expressions:

$$\boldsymbol{\varepsilon}_{12} = \mathbf{T}\,\boldsymbol{\varepsilon} \qquad \boldsymbol{\sigma} = \mathbf{T}^T\boldsymbol{\sigma}_{12} \qquad (4)a,b$$

The rotation matrix $\mathbf{T}$ is defined by

$$\mathbf{T} = \begin{bmatrix} \sin^2\theta & \cos^2\theta & \sin\theta\cos\theta \\ \cos^2\theta & \sin^2\theta & -\sin\theta\cos\theta \\ 2\sin\theta\cos\theta & -2\sin\theta\cos\theta & \cos^2\theta - \sin^2\theta \end{bmatrix} \quad (5)$$

The solution of nonlinear problems demands linearized incremental equations. An increment of stress $\Delta\boldsymbol{\sigma}_{12}$ is related to an increment of strain $\Delta\boldsymbol{\varepsilon}_{12}$ according to

$$\Delta\boldsymbol{\sigma}_{12} = \mathbf{E}_{12}\,\Delta\boldsymbol{\varepsilon}_{12} \quad (6)$$

$$\mathbf{E}_{12} = \begin{bmatrix} E_{11} & E_{12} & 0 \\ E_{21} & E_{22} & 0 \\ 0 & 0 & G_{12} \end{bmatrix} \quad (7)$$

Equations (3) to (5) define a rotating orthotropic model steered by the principal direction $\theta$. Since the constitutive functions (3) are sufficient to relate any state of strain to the corresponding stresses, they must also be sufficient to define the tangent constitutive matrix $\mathbf{E}_{12}$, which is not necessarily symmetric. The elasticity moduli of concrete $E_{ij}$ $(i, j = 1, 2, 3)$ and the tangent shear modulus $G_{12}$ are given by

$$E_{ij} = \partial\sigma_i / \partial\varepsilon_j \quad (8)$$

$$G_{12} = [\sigma_2 - \sigma_1] / [2(\varepsilon_2 - \varepsilon_1)] \quad (9)$$

Derivatives (8) are established according to the constitutive equations (3). Expression (9) is proved for the general three-dimensional case by Schulz and Santisi d'Avila (2010). The relation between stress and strain increments in the $xz$ coordinate system is expressed by

$$\Delta\boldsymbol{\sigma} = \mathbf{E}\,\Delta\boldsymbol{\varepsilon} \quad (10)$$

where the tangent constitutive matrix of concrete, $\mathbf{E}$, is $\mathbf{E} = \mathbf{T}^T \mathbf{E}_{12}\, \mathbf{T}$.

The constitutive equations of the longitudinal and transverse reinforcements are $\sigma_{sx} = \sigma_{sx}(\varepsilon_{sx})$ and $\sigma_{sw} = \sigma_{sw}(\varepsilon_{sw})$. The following equation establishes the relation between the increment of the steel stress vector $\Delta\boldsymbol{\sigma}_s$ and the strain increment $\Delta\boldsymbol{\varepsilon}$:

$$\Delta\boldsymbol{\sigma}_s = \mathbf{E}_s\,\Delta\boldsymbol{\varepsilon} \quad (11)$$

The tangent constitutive matrix of reinforcement $\mathbf{E}_s$ is defined by

$$\mathbf{E}_s = \begin{bmatrix} E_{sx} & 0 & 0 \\ 0 & E_{sw}\cos^2 w & 0 \\ 0 & 0 & 0 \end{bmatrix} \quad (12)$$

$E_{sx}$ and $E_{sw}$ are the tangent stiffness of longitudinal and transverse bars. Using (1), (10) and (11), the increment $\Delta\mathbf{f}$ of total internal forces per unit length in reinforced concrete is

$$\Delta\mathbf{f} = \mathbf{C}\,\Delta\boldsymbol{\varepsilon} \quad (13)$$

where the tangent constitutive matrix $\mathbf{C}$ of the reinforced concrete element in $xz$ coordinates is expressed by $\mathbf{C} = b\,\mathbf{E} + \mathbf{a}_s\,\mathbf{E}_s$.

## 3 Reinforced concrete cross-section

According to hypothesis 7, the applied loads per unit length are assumed to be zero. The differential equilibrium equations of the reinforced concrete element are expressed by

$$f_x' + f_{xz}^* = 0 \tag{14}$$

$$f_{xz}' + f_z^* = 0 \tag{15}$$

In equations (14), (15) and in the following, the prime $(\ )'$ and the star $(\ )^*$ denote, respectively, partial derivatives of a function with respect to $x$ $(\partial/\partial x)$ and $z$ $(\partial/\partial z)$. Hypothesis 4 $(f_z \cong 0)$ and equation (15) yield $f_z' \cong 0$, $f_z^* \cong 0$ and $f_{xz}' \cong 0$.

Expression (13) can be expanded according to the following expression:

$$\begin{bmatrix} \Delta f_x \\ \Delta f_z \\ \Delta f_{xz} \end{bmatrix} = \begin{bmatrix} c_{11} & c_{12} & c_{13} \\ c_{21} & c_{22} & c_{23} \\ c_{31} & c_{32} & c_{33} \end{bmatrix} \begin{bmatrix} \Delta \varepsilon_x \\ \Delta \varepsilon_z \\ \Delta \varepsilon_{xz} \end{bmatrix} \tag{16}$$

When $\Delta f_z = \Delta f_{xz} = 0$, expression (16) reduces to

$$\Delta f_x = D\, b\, \Delta \varepsilon_x \tag{17}$$

where the equivalent uniaxial elasticity modulus $D$ of the reinforced concrete element is

$$D = \frac{c_{13}\, c_{22}\, c_{31} - c_{12}\, c_{23}\, c_{31} - c_{13}\, c_{21}\, c_{32} + c_{11}\, c_{23}\, c_{32} + c_{12}\, c_{21}\, c_{33} - c_{11}\, c_{22}\, c_{33}}{b\left(c_{23}\, c_{32} - c_{22}\, c_{33}\right)} \tag{18}$$

Equation (17) define $f_x'$ as

$$f_x' = D\, b\, \varepsilon_x' \tag{19}$$

According to hypothesis 7, the shear flow $f_{xz}$ is zero at the boundary $z = z_A$. Integrating (14) and applying (19) yields

$$f_{xz}(z) = -\int_{z_A}^{z} f_x'\, \mathrm{d}z = -\int_{z_A}^{z} D\, b\, \varepsilon_x'\, \mathrm{d}z \tag{20}$$

As stated in hypothesis 6, plane sections remain plane after deformation. The longitudinal strain $\varepsilon_x$ is linearly interpolated by

$$\varepsilon_x = e_x + k_x\, z = \mathbf{p}^T\, \mathbf{e} \tag{21}$$

where $e_x$ is the longitudinal strain at $z = 0$ and $k_x$ is the curvature of the cross-section. The position vector $\mathbf{p}$ and the vector of generalized strains $\mathbf{e}$ are defined by $\mathbf{p} = \begin{bmatrix} 1 & z \end{bmatrix}^T$ and $\mathbf{e} = \begin{bmatrix} e_x & k_x \end{bmatrix}^T$. The derivative of (21) with respect to $x$ gives

$$\varepsilon_x' = \mathbf{p}^T\, \mathbf{e}' \tag{22}$$

where $\mathbf{e}'$ is the vector of the derivatives of the generalized strains, expressed by $\mathbf{e}' = \begin{bmatrix} e_x' & k_x' \end{bmatrix}^T$. Substituting (22) in (20) yields

$$f_{xz}(z) = \mathbf{S}^T(z)\,\mathbf{e}' \tag{23}$$

where $\mathbf{S}(z)$ is a vector defined by

$$\mathbf{S}(z) = \begin{bmatrix} S_1(z) & S_z(z) \end{bmatrix}^T = -\int_{z_A}^{z} \mathbf{p}\, D\, b\, dz \tag{24}$$

An equivalent cross-section is constructed by multiplying, at each level $z$, the reinforced concrete section $b\,dz$ and the equivalent uniaxial modulus $D$ (**Fig. 3**). The parameters $S_1(z)$ and $S_z(z)$ represent the equivalent area of the cross-section, between $z_A$ and $z$, and the corresponding first moment with respect to the $y$-axis, both with opposite signs.

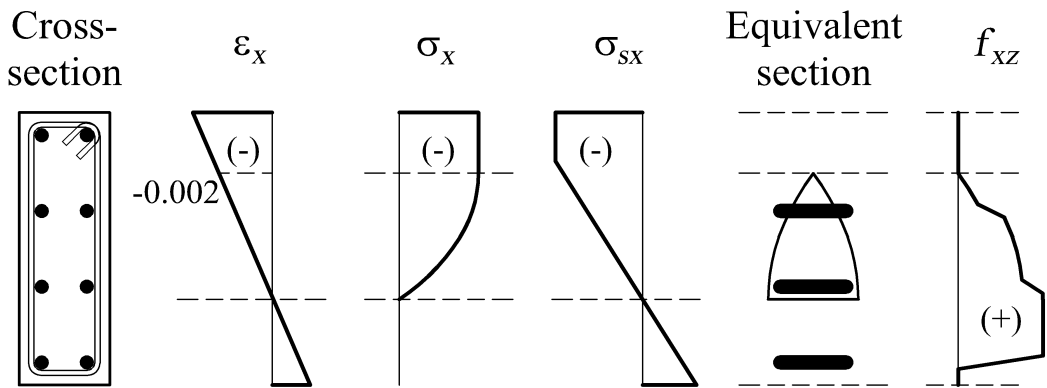

**Fig. 3**   The Equivalent Section

The vector $\mathbf{F}$ of the generalized stresses and its derivative are respectively expressed by

$$\mathbf{F} = \begin{bmatrix} N & M \end{bmatrix}^T = \int_{z_A}^{z_B} \mathbf{p}\, f_x\, dz \tag{25}$$

$$\mathbf{F}' = \begin{bmatrix} 0 & V \end{bmatrix}^T = \int_{z_A}^{z_B} \mathbf{p}\, f_x'\, dz \tag{26}$$

The derivative of the normal force $N$ is zero, as stated in hypothesis 7. The derivative of the bending moment $M$ is the shear force $V$. Substituting (19) and (22) in equation (26) yields

$$\mathbf{F}' = \mathbf{K}\,\mathbf{e}' \tag{27}$$

The stiffness matrix $\mathbf{K}$ of the cross-section is defined by

$$\mathbf{K} = \int_{z_A}^{z_B} \mathbf{p}\, D\, \mathbf{p}^T b\, dz = \begin{bmatrix} I_{11} & I_{1z} \\ I_{z1} & I_{zz} \end{bmatrix} \tag{28}$$

where the stiffness parameters $I_{mn}$ are respectively equal to $I_{mn} = \int_{z_A}^{z_B} m\, n\, D\, b\, dz$. The shear flow $f_{xz}$ and derivatives of the generalized strains $\mathbf{e}'$ are determined using (23) and (27). It is possible to select a principal coordinate system such that the stiffness matrix $\mathbf{K}$ becomes diagonal. Specifying the coordinate axis $\bar{z} = z - z_G$, such that $z_G = I_{1z}/I_{11}$ is the $z$-coordinate of the barycenter, yields $\bar{e}' = 0$ and $\bar{k}' = V/I_{\bar{z}\bar{z}}$. Equations (23) and (27) simplify to the Jouravski formula

$$f_{xz}(\bar{z}) = V\, S_{\bar{z}}(\bar{z}) / I_{\bar{z}\bar{z}} \tag{29}$$

$S_{\bar{z}}(\bar{z})$ is the first moment, with an opposite sign, of the equivalent area between $\bar{z}_A$ and $\bar{z}$. The principal moment of inertia of the "equivalent section" is $I_{\bar{z}\bar{z}}$. The "equivalent section" helps not only to determine but also to understand the shear flow in cross-sections of nonlinear materials. The shear stresses remain constant when the tangent modulus $D$ is equal to zero. The shear flow in the extreme fibers is equal to zero when plastic materials reach their maximum capacity to carry longitudinal stresses.

## 4 Material behaviour

The constitutive model A suggested by Vecchio and Collins (1993) is adopted for implementation. In model A, the softening effect in tension-compression state is expressed as a function of the ratio $\varepsilon_1/\varepsilon_2$, where $\varepsilon_1$ and $\varepsilon_2$ are the principal tensile and compressive strains, respectively. The tension-stiffening effect is represented as proposed by Polak and Vecchio (1993). Shear slip at crack surfaces, Poisson's ratio and other secondary effects are not considered. Other details are discussed by Schulz and Santisi d'Avila (2010). The terms $E_{ij}$, defined in (8), are determined by the following expression:

$$E_{ij} = \frac{\partial \sigma_i}{\partial \varepsilon_j} + \frac{\partial \sigma_i}{\partial \beta}\left(\frac{\partial \beta}{\partial \varepsilon_1}\frac{\partial \varepsilon_1}{\partial \varepsilon_j} + \frac{\partial \beta}{\partial \varepsilon_2}\frac{\partial \varepsilon_2}{\partial \varepsilon_j}\right) \tag{30}$$

where $\beta(\varepsilon_1, \varepsilon_2)$ is the tension-softening coefficient. The derivatives in (30), are analytically presented by Santisi d'Avila (2008).

## 5 Numerical procedure

The computational solution is implemented by dividing the cross-section, along the height, into a sufficient number of plane elements. The following numerical procedure yields strains and stresses for a given set of internal forces:

1. An iteration is started considering a shear flow $f_{xz}$ along the height and a generalized strain vector $\mathbf{e}$ (21). Both of the initial approximations can be equal to zero.

2. At each layer, the terms $\varepsilon_x(z)$ (21), $f_z(z) \cong 0$ and $f_{xz}(z)$ are known. The terms $f_x(z)$, $\varepsilon_z(z)$ and $\gamma_{xz}(z)$ are determined using a secondary iterative process based on (16) which final step yields the equivalent uniaxial modulus $D$ (18).

3. The process stops when the residual $\Delta\mathbf{F}$ between the applied and the resisting internal forces $\mathbf{F}$ (25) is considered relatively small. The stiffness matrix $\mathbf{K}$ (28) and the derivative of the generalized strains $\mathbf{e}'$ (27) yield a new approximation of the shear flow $f_{xz}$ (23). Solving $\Delta\mathbf{F} = \mathbf{K}\,\Delta\mathbf{e}$ yields the strain increment $\Delta\mathbf{e}$ and a new approximation of the generalized strain vector $\mathbf{e}$, restarting the main iterative process. The simultaneous update of the shear flow and generalized strains proves to be numerically efficient, although each estimation assumes that the other parameter is restrained.

## 6  Comparison with test data reported in the literature

The proposed theory is verified through test results of beams with two symmetric point loads reported in the literature. ET1 to ET4 is a group of reinforced concrete beams tested by Leonhardt and Walter (1962), demonstrating that cross-section shape influences the shear capacity. Material and geometrical conditions are the same, but web widths vary. Although other secondary effects are not taken into account, the influence of the web thickness on stirrup stresses is detected and conservatively represented.

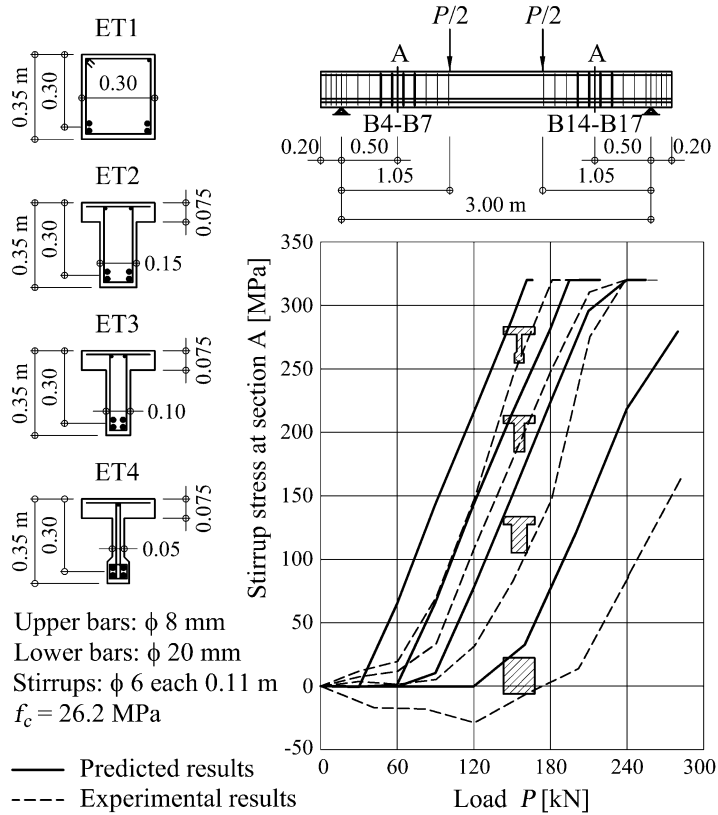

**Fig. 4**  Beams ET1 to ET4

The second group of specimens is selected from prestressed beams tested by Elzanaty, Nilson and Slate (1985). The I-beams (CW) and T-beams (CI) present concrete strengths varying between 40 to 80 MPa (**Fig. 5**). The beams are divided into two groups: beams with stirrups and beams without stirrups. Two types of reinforcement are used: deformed bars $\left( f_y = 434\,\mathrm{MPa} \right)$ and smooth wires of 6.4mm diameter $\left( f_{2‰} \approx 380\,\mathrm{MPa} \right)$. Each beam contains four 7-wire low relaxation strands, of 0.5 or 0.6 inch nominal diameter. The corresponding stress at 1% strain is approximately 1800 MPa. The residual strains $\varepsilon_{sx0}$ of the strands are evaluated based on the effective prestress force $F_p$ applied in each beam, as given by Elzanaty et al. (1985) by taking into account all losses that take place from the prestressing operation to testing. $M_u$ and $V_u$ indicate the ultimate bending moment and shear force. The shear reinforcement parameter $\chi$ is defined by $\chi = 0.9\, d\, f_{yw}\, A_{sw} / V_u$. The predicted ultimate loads are compared to the experimental data (**Tab. 1**) and demonstrate good correlation.

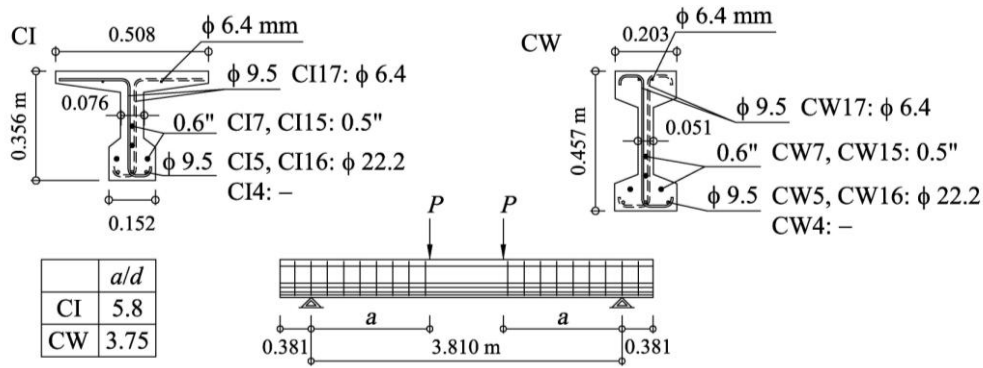

**Fig. 5** Geometry of CI and CW prestressed beams

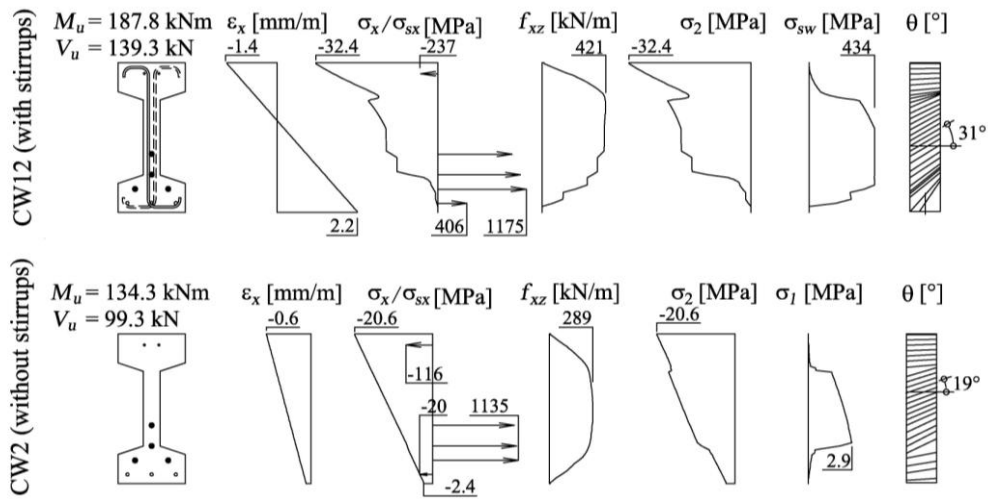

**Fig. 6** Ultimate analysis

**Tab. 1** Predicted and observed capacity

Beams with stirrups

| Sp. | $f_c$ [MPa] | $V_u$ [kN] | $\chi$ [%] | $\dfrac{V_{u\,PREDICTED}}{V_u}$ [%] |
|---|---|---|---|---|
| CI10 | 73.1 | 141 | 49 | 91-95 |
| CI11 | 55.8 | 127 | 54 | 98-102 |
| CI12 | 40.0 | 122 | 57 | 98-102 |
| CI13 | 72.4 | 155 | 45 | 89-92 |
| CI14 | 73.8 | 165 | 42 | 100-102 |
| CI15 | 70.3 | 121 | 58 | 94-100 |
| CI16 | 73.1 | 163 | 46 | 107-110 |
| CI17 | 69.6 | 129 | 21 | 78-87 |
| CW10 | 73.1 | 173 | 45 | 86-92 |
| CW11 | 55.8 | 157 | 50 | 92-97 |
| CW12 | 40.0 | 141 | 55 | 98-103 |
| CW13 | 72.4 | 182 | 43 | 92-97 |
| CW14 | 73.8 | 188 | 41 | 102-106 |
| CW15 | 70.3 | 150 | 52 | 97-101 |
| CW16 | 73.1 | 187 | 44 | 99-104 |
| CW17 | 69.6 | 142 | 21 | 73-84 |

Beams without stirrups

| Sp. | $f_c$ [MPa] | $V_u$ [kN] | $\dfrac{V_{u\,PREDICTED}}{V_u}$ [%] |
|---|---|---|---|
| CI2 | 76.5 | 116 | 74-79 |
| CI4 | 78.6 | 109 | 82-86 |
| CI5 | 77.9 | 120 | 66-73 |
| CI6 | 77.9 | 89 | 74-88 |
| CI7 | 77.6 | 81 | 76-94 |
| CI8 | 41.4 | 85 | 76-81 |
| CI9 | 61.0 | 87 | 74-86 |
| CW2 | 76.5 | 125 | 79-80 |
| CW4 | 78.6 | 127 | 80-81 |
| CW5 | 77.9 | 124 | 76-78 |
| CW6 | 77.9 | 112 | 75-77 |
| CW7 | 77.6 | 106 | 78-80 |
| CW8 | 41.4 | 90 | 81-83 |
| CW9 | 61.0 | 101 | 78-80 |

## 7 Conclusions

This research presents a procedure for the analysis of reinforced concrete beams considering the simultaneous effect of axial load, bending moment. The shear flow is determined by applying the Jouravski formula to the "equivalent section". The whole set of equilibrium, compatibility and constitutive equations is satisfied, the stiffness derivatives are explicitly calculated and the algorithms show good convergence. The validation of the proposed model is undertaken by comparison with experimental results obtained by other researchers. The examples confirm that the theory is able to accurately model the behaviour or reinforced concrete beams. The "equivalent section" helps not only to determine but also to understand the shear flow in cross-sections of nonlinear materials. The shear stresses remain constant when the tangent modulus $D$ is equal to zero. The shear flow in the extreme fibers is equal to zero when plastic materials reach their maximum capacity to carry longitudinal stresses. The "equivalent section" is a powerful and practical tool for advanced reinforced concrete design.

**Prof. Mauro Schulz**
✉ Universidade Federal Fluminense
Departamento de Engenharia Civil
Rua Passo da Patria 156
24210-240 Niteroi, RJ, Brazil
☎ +55 21 26 29 54 48
📠 +55 21 81 69 80 67
☺ mschulz@attglobal.net
**URL**

**Maria Paola Santisi d'Avila, Ph.D.**
✉ École Polytechnique
Laboratoire de Mécanique des Solides
Route de Saclay
91120 Palaiseau, IDF, France
☎ +33 69 33 57 71
📠 +33 69 33 30 26
☺ mpaolasantisi@gmail.com
**URL**